\documentclass[aps,prl,twocolumn,superscriptaddress,floatfix]{revtex4} 
\usepackage{epsf} 
\usepackage{epsfig} 
 
\def\utfit{{\bf{U}}\kern-.24em{\bf{T}}\kern-.21em{\it{fit}}\@}
\def\bc{\begin{center}}
\def\ec{\end{center}}
\def\be{\begin{equation}}
\def\ee{\end{equation}}
\def\bea{\begin{eqnarray}}
\def\eea{\end{eqnarray}}

\def\lsim{\mathrel{\rlap{\lower4pt\hbox{\hskip1pt$\sim$}}
    \raise1pt\hbox{$<$}}}                
\def\gsim{\mathrel{\rlap{\lower4pt\hbox{\hskip1pt$\sim$}}
    \raise1pt\hbox{$>$}}}                
\begin{document}

\preprint{RM3-TH/05-5}
\preprint{TUM-HEP-595/05}

\title{THE EFFECT OF PENGUINS IN THE $B_d \to J/ \psi K^0$ CP
    ASYMMETRY}

\author{M.~Ciuchini}
\affiliation{Dip. di Fisica, Univ. di Roma Tre 
and INFN, Sezione di Roma Tre, Via della Vasca Navale 84, I-00146   
Roma, Italy}
\author{M.~Pierini}
\affiliation{Department of Physics, University of Wisconsin, Madison,
  WI 53706, USA} 
\author{L.~Silvestrini}
\affiliation{Physik-Department, Technische Universit{\"a}t
  M{\"u}nchen, D-85748 Garching, Germany.}
\altaffiliation{Dip. di Fisica, Univ. ``La Sapienza''  and INFN,
  Sezione di Roma, P.le A. Moro, I-00185 Rome, Italy.} 

\begin{abstract}
  Performing a fit to the available experimental
  data, we quantify the effect of long-distance contributions from
  penguin contractions in $B^0 \to J/ \psi K^0$ decays. We estimate
  the deviation of the measured $\mathcal{S}_\mathrm{CP}$ term of
  the time-dependent CP asymmetry from $\sin2\beta$ induced by these
  contributions and by the penguin operators. We find $\Delta
  \mathcal{S} \equiv \mathcal{S}_\mathrm{CP}(J/\psi K) - \sin 2
  \beta = 0.000 \pm 0.017$ ($[-0.035,0.033]\; @ \, 95\%$
  probability), an uncertainty much larger than previous estimates
  and comparable to the present systematic error quoted by the
  experiments at the $B$-factories.
\end{abstract}

\maketitle

The measurement of the phase of the $B_d - \bar B_d$ mixing amplitude,
given by twice the angle $\beta$ of the Unitarity Triangle (UT) in the
Standard Model (SM), is one of the main successes of $B$-factories,
and a crucial ingredient to test the SM and to look for new physics.
The golden mode for this measurement is given by $B^0 \to J/\psi K^0$
decays~\cite{sanda}. These modes give a value of $\sin 2\beta$ which
is considered practically free of theoretical uncertainties and thus
serves as a benchmark for indirect searches for new physics. Indeed,
new physics can reveal itself by comparing different observables --
which all determine $\sin 2\beta$ in the SM -- to the reference value
from the $J/\psi K^0$ modes.  For instance, $\sin 2\beta$ can be
extracted from the UT fit or from $b\to s$ penguin-dominated modes
such as $B\to \phi K_s$ or $B\to \eta^\prime K_s$.  Actually, possible
hints of a discrepancy are being seen in both
cases~\cite{ckmfit,utfit}.

Impressive progress has been recently achieved at the $B$-factories in the
measurement of the coefficient $\mathcal{S}_\mathrm{CP}$ of the time-dependent
CP asymmetry in $B^0\to J/\psi K^0$ decays. The experimental error on
$\mathcal{S}_\mathrm{CP}$ has been pushed down to $\pm 0.028$ (statistical) $\pm
0.020$ (systematic)~\cite{exp-sin2b}. On the theoretical side, previous
estimates of the uncertainty in the extraction of $\sin 2\beta$ from
$\mathcal{S}_\mathrm{CP}$ gave results below $10^{-3}$ (for a recent study, see
ref.~\cite{mannel}) and therefore completely negligible. In this paper, we
reanalyze this issue with a new approach, described in detail below, obtaining
a substantially larger uncertainty comparable to the present experimental
systematic error.

The decays of neutral $B$ mesons into $J/\psi K^0$ final states are
dominated by a tree-level amplitude proportional to $V_{cb}V^*_{cs}$.
Assuming the absence of additional contributions with different weak
phases, it is possible to extract the value of $\sin 2\beta$ from
the coefficient $\mathcal{S}_\mathrm{CP}$ of the time-dependent CP
asymmetry in these decays. As already mentioned, the identification of
$\mathcal{S}_\mathrm{CP}(J/ \psi K_{S/L})$ with $\sin 2\beta$ is
affected by a theoretical uncertainty, coming from the presence of
additional contributions having a different weak phase and possibly a
relative strong phase with respect to the dominant
contribution~\cite{Grinstein}.  Using the OPE, we write the expression
of the decay amplitudes arranging all the contractions of effective
operators into renormalization group invariant 
parameters~\cite{buras-silv}. In this way, we have
\begin{eqnarray}
A(B^0 \to J/\psi K^0) &=& V^*_{cb}V_{cs}(E_2-P_2) \nonumber\\
&&+V^*_{ub}V_{us}(P_2^\mathrm{GIM}-P_2)\,, 
\label{eq:amp}
\end{eqnarray}
where $E_2$ represents the dominant tree contribution and the other
terms are penguin corrections. Although three parameters ($E_2$, $P_2$
and $P_2^\mathrm{GIM}$) enter the amplitude, for the purpose of this
paper they can be treated as two effective parameters $E_2-P_2$ and
$P_2^\mathrm{GIM}-P_2$. Neglecting the doubly Cabibbo-suppressed
combination $P_2^\mathrm{GIM}-P_2$, a penguin pollution could come
from $P_2$. Even though this contribution might have an impact on the
branching ratio, it certainly does not affect the CP asymmetry, since
the two amplitudes carry the same weak phase.  Conversely, because of
the weak phase of $V_{ub}$, $P_2^\mathrm{GIM}-P_2$ might produce an
effect on $\mathcal{S}_\mathrm{CP}$ and $\mathcal{C}_\mathrm{CP}$,
although the impact on the branching ratio is expected to be very
small.

Being doubly Cabibbo suppressed, the value of $P_2^\mathrm{GIM}-P_2$
is hardly determined from $B \to J/\psi K$ decays alone. Therefore one
needs to extract the range of this parameter from a different decay in
order to study the impact of such a subdominant effect on $\sin 2
\beta$. Indeed, the induced uncertainty on $\mathcal{S}_\mathrm{CP}$
increases with the upper bound of this range. It is then of the utmost
importance to quantify this upper bound in a reliable way. Previous
detailed discussions of the uncertainty $\Delta \mathcal{S} \equiv
\mathcal{S}_\mathrm{CP}(J/\psi K) - \sin 2 \beta$ have estimated the
effect of $P_2^\mathrm{GIM}-P_2$ using the BSS mechanism~\cite{BSS},
recently supported by QCD factorization, to express penguin
contractions in terms of local four-fermion operators~\cite{mannel}.
However, QCD factorization holds only formally for this
channel \footnote{Subleading terms are only suppressed as
  $\Lambda_{QCD}/(m_b \alpha_s)$, so that the suppression is marginal
  for the actual value of $m_b$ \cite{BBNSlong}.}. Clearly, the
importance of this measurement for testing the SM and looking for new
physics calls for a more general assessment of the theoretical
uncertainty. In the present work, we aim at providing a
model-independent estimate of $\Delta \mathcal{S}$.

To fulfill our task, we proceed in three steps:
i) Neglecting $P_2^\mathrm{GIM}-P_2$, we extract the absolute value
  of $E_2-P_2$, using the experimental value of the branching ratio.
ii) We extract $\vert E_2-P_2 \vert$, $\vert P_2^\mathrm{GIM}-P_2
  \vert$ and the relative strong phase $\delta_P$ from a fit to the
  SU(3)-related (up to the assumption discussed below) channel $B^0\to
  J/ \psi \pi^0$. In this decay mode, $P_2^\mathrm{GIM}-P_2$ is not
  doubly Cabibbo suppressed and can be determined with good accuracy.
  At the same time, we can compare the value of $E_2-P_2$ obtained in
  the two channels to test the SU(3) invariance and the additional
  assumption.  We can then take the range of $P_2^\mathrm{GIM}-P_2$
  from this fit (at $99.9 \%$ probability) as a reliable estimate of
  the range to be used in $B^0\to J/ \psi K^0$.
iii) We repeat the first step, varying $P_2^\mathrm{GIM}-P_2$ in
  the range obtained in the second step. In this way, we get the
  distribution of $\mathcal{S}_\mathrm{CP}$, to be compared with the input
  $\sin2\beta$ to obtain $\Delta \mathcal{S}$.

Let us provide some details about the second step. Using the same
formalism of Eq.~(\ref{eq:amp}) we can write the decay amplitude of
$B^0\to J/ \psi \pi^0$ as:
\begin{eqnarray}
A(B^0 \to J/\psi \pi^0) &=& 
V^*_{cb}V_{cd}(E_2-P_2) \nonumber \\
&& + V^*_{ub}V_{ud}(P_2^\mathrm{GIM}-P_2)\,,
\label{eq:amppi0}
\end{eqnarray}
where all the combinations of CKM elements now are of the same order
of magnitude and the additional (OZI-suppressed) contribution of the
emission-annihilation $\textit{EA}_2$ parameter has been
ignored \footnote{This approximation can be tested using BR$(B^0 \to
  D^0 \phi)$ which is proportional to $\textit{EA}_2$.}. Even though
the SU(3) symmetry is not exact (so that assuming the parameters
to be the same in the two fits would require a difficult estimate of
the associated error), we think that SU(3) is good enough to give us a
reasonable estimate of the allowed range of $|P_2^\mathrm{GIM}-P_2|$.

In the three fits, we use as input the
determination of the CKM matrix obtained by the \utfit\ Collaboration
discarding the bound on $\bar \rho$ and $\bar \eta$  from $B^0 \to J/
\psi K^0$~\cite{utfit}. To give a reference normalization 
factor for all the results, we use the value of $E_2$, computed using naive
factorization.
All the inputs used in the fit are summarized in
Tab.~\ref{tab:inputs}.  We assume flat distributions for $F^{B \to
  \pi}$ and for $F^{B \to K}/F^{B \to \pi}$ in the ranges specified
\cite{latticeQCD}.

\begin{table}[tb]
\caption{Input values used in the analysis. All dimensionful
  quantities are given in GeV.\label{tab:inputs}}
\begin{ruledtabular}
\begin{tabular}{cccc}
  $F^{B \to \pi}$  & $0.27 \pm 0.08$ & $F^{B \to K}/F^{B \to \pi}$ & $1.2 \pm 0.1$ \\ 
  $f_{J/\psi}$   & $0.131$ & $m_B$       & $5.2794$ \\
  $\bar \rho $        & $0.207 \pm 0.038$     & $\bar \eta$ & $0.341 \pm 0.023$ \\ 
  $A$         & $0.86 \pm 0.04$       & $\lambda$         & $0.2258 \pm 0.0014$ \\ 
  $G_F$       & $1.166\cdot 10^{-5}$  & $\alpha_{em}$      & $ 1/129$ \\
\end{tabular}
\end{ruledtabular}
\end{table}

Using the experimental value of BR$(B^0\to J/\psi K^0)$, we bound the
absolute value of $E_2-P_2$ \footnote{We can redefine the overall
  phase in such a way that this contribution is real.}.  Using the
statistical method of \utfit{}~\cite{utfit2000}, we assign a flat
\textit{a-priori} distribution to the absolute value $|E_2-P_2|$ in a
range large enough to fully include the region where the
\textit{a-posteriori} distribution is non-vanishing. In this way, we
reproduce the experimental value of the branching ratio with an
indication of a significant effect of nonfactorizable corrections in
$|E_2-P_2|$, as already noted in \cite{melic}. We obtain $|E_2-P_2| =
1.44 \pm 0.05$.  Notice that, in the single-amplitude approximation
used in this first step, the predicted $\mathcal{C}_\mathrm{CP}$ is
exactly vanishing while $\mathcal{S}_\mathrm{CP}$ is, as expected,
equal to the input value for $\sin 2\beta$ ($\mathcal{S}_\mathrm{CP} =
0.729 \pm 0.042$).

\begin{figure}[tb]	
  \includegraphics[width=0.3\textwidth]{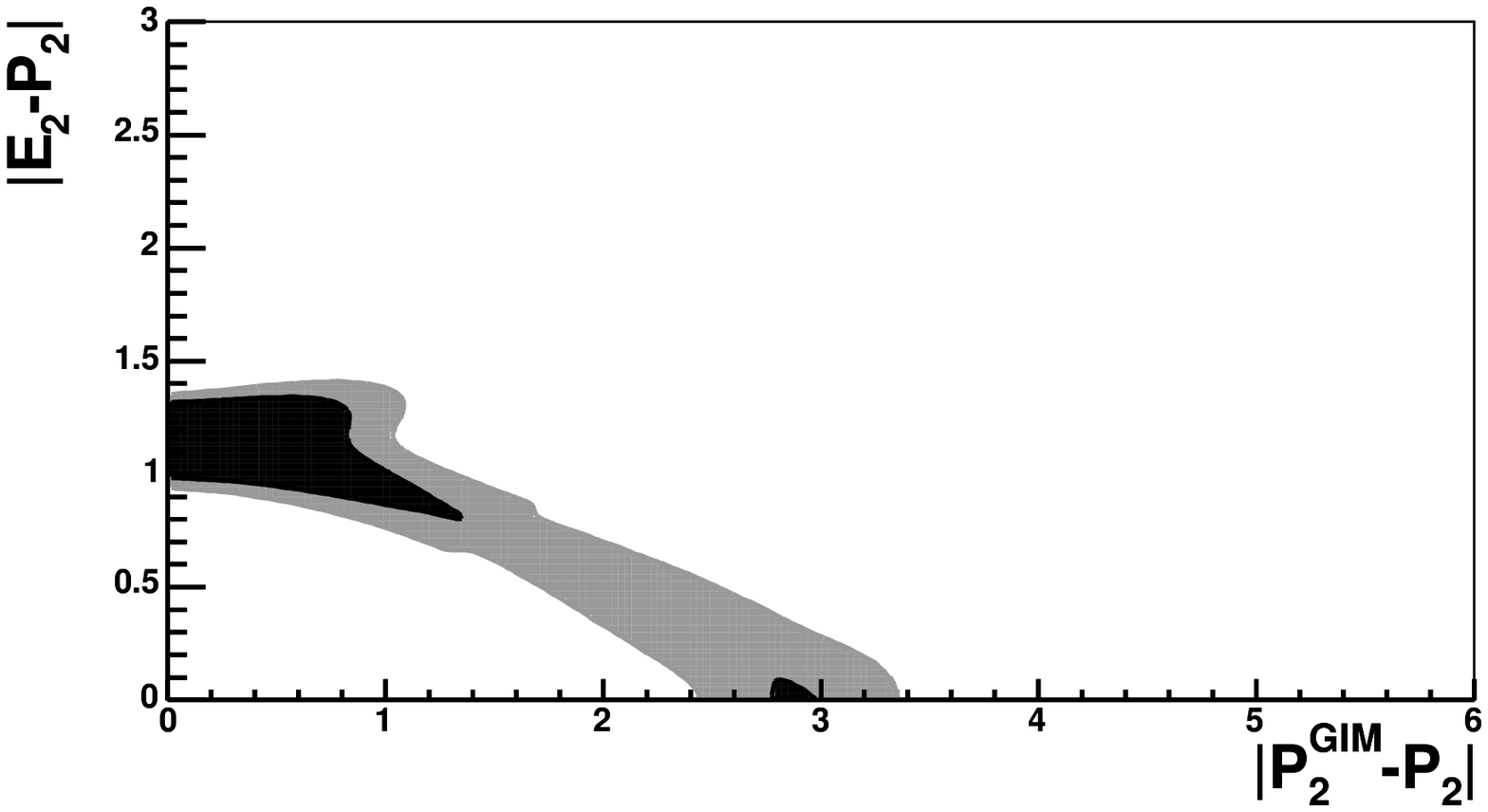} 
  \caption{Correlation between the
    hadronic parameters $\vert E_2-P_2\vert$ and $\vert
    P_2^\mathrm{GIM}-P_2\vert$, as obtained from the fit to
    $\bar B^0 \to J/ \psi \pi^0$.}
  \label{fig:results2}
\end{figure}

\begin{figure*}[tb]	
    \includegraphics[width=0.28\textwidth]{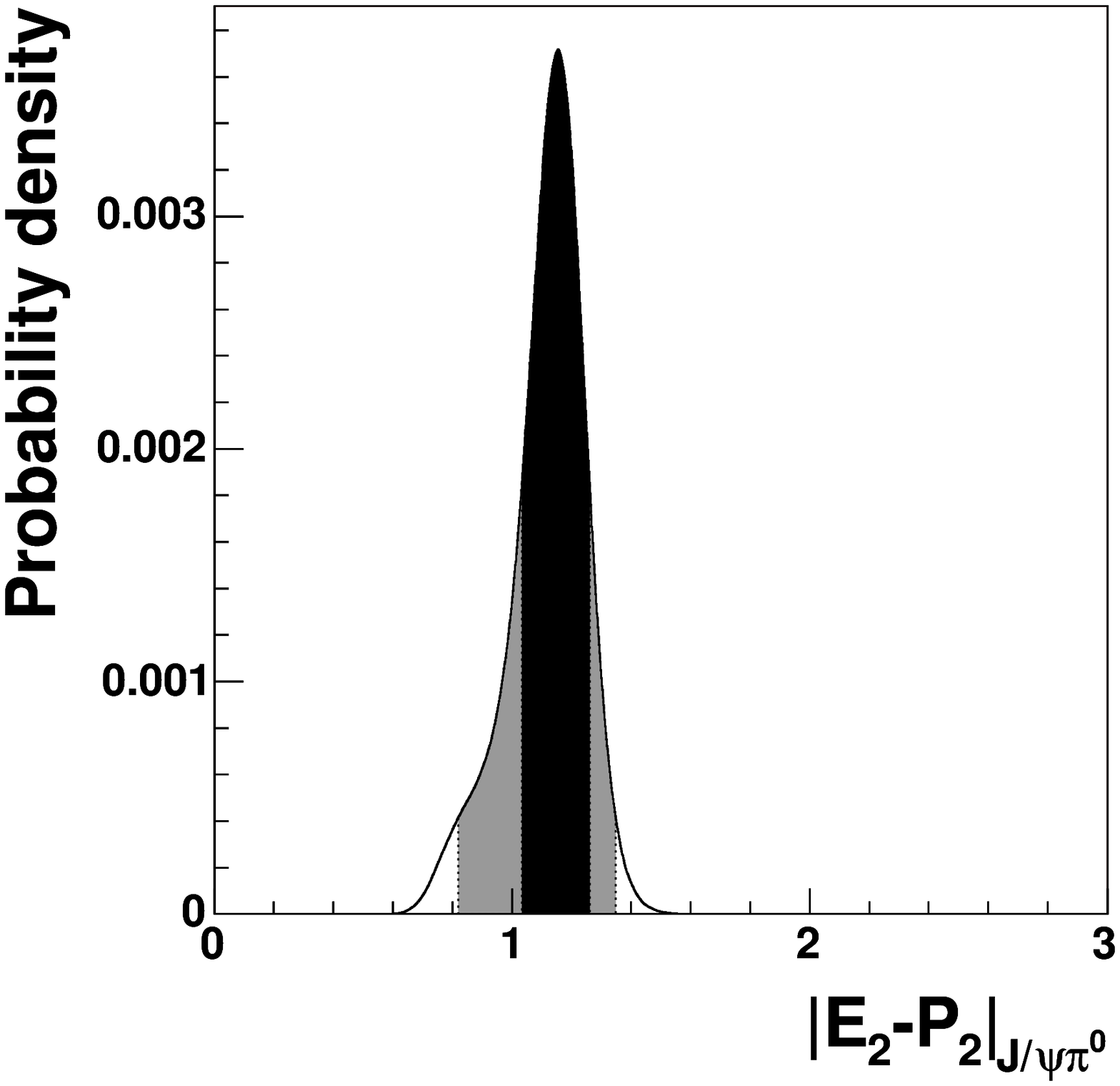} 
    \includegraphics[width=0.28\textwidth]{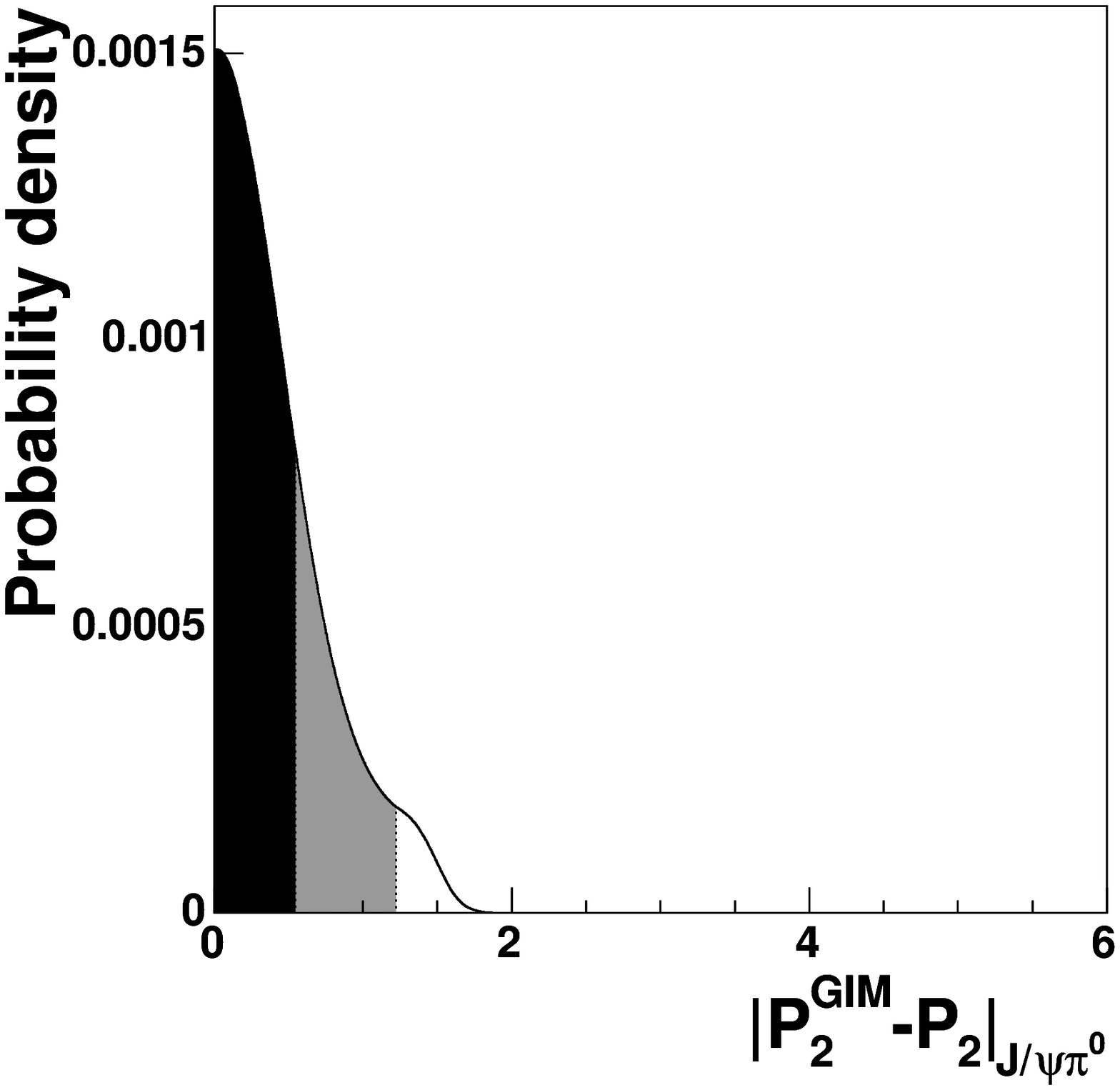}
    \includegraphics[width=0.28\textwidth]{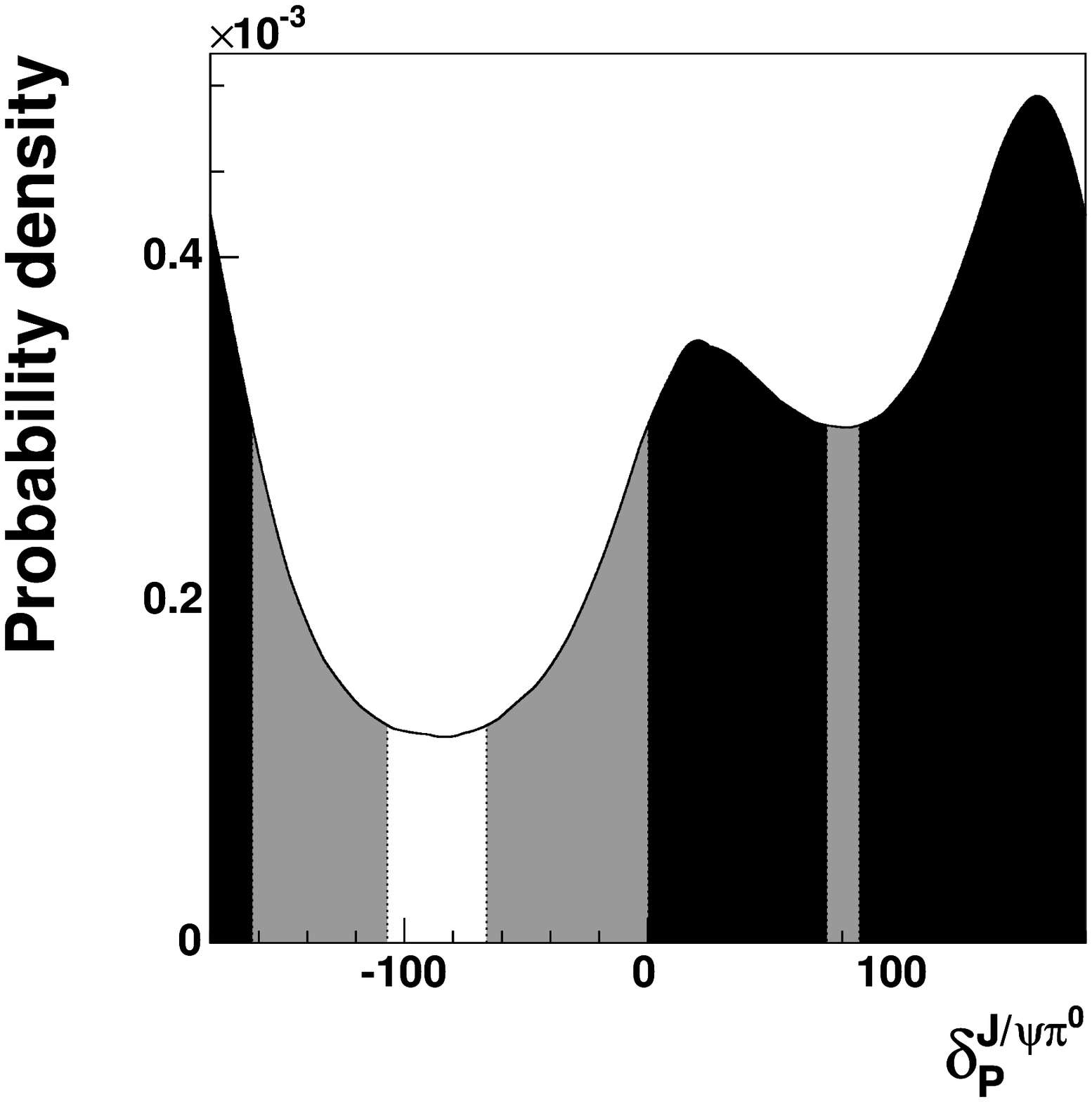} 
\caption{Output distributions of hadronic parameters $\vert
  E_2-P_2\vert$ (top left), $\vert P_2^\mathrm{GIM}-P_2\vert$ (top right) and
  $\delta_{P}$ (bottom), as obtained from the fit to $\bar B^0 \to J/
  \psi \pi^0$ with the cut $\vert P_2^\mathrm{GIM}-P_2\vert < 2 \vert
  E_2-P_2\vert$ (see the text for details).}
\label{fig:P2cut}
\end{figure*}

We now extract $P_2^\mathrm{GIM} -P_2$ from $B^0 \to J/\psi \pi^0$.
For this fit, we use the same approach but we retain in the amplitude
$\vert E_2-P_2\vert $, $\vert P_2^\mathrm{GIM}-P_2\vert$ and the
relative strong phase $\delta_P$.  Together with the experimental
information from the branching ratio and $\mathcal{C}_\mathrm{CP}$, we
impose the constraint coming from
$\mathcal{S}_\mathrm{CP}$~\cite{jpsipi0exp}. We allow $\vert
E_2-P_2\vert $ and $\vert P_2^\mathrm{GIM} -P_2\vert$ to vary in a
range larger than the support of the output distributions, and
$\delta_P \in [-\pi,\pi]$.  The results are given in
Tab.~\ref{tab:results2}.
\begin{table}[tb]
\caption{Results of the fit of $\bar B^0 \to J/ \psi \pi^0$ (see the
  text for details).}
\label{tab:results2}
\begin{ruledtabular}
\begin{tabular}{cccc}
  $\mathcal{C}_\mathrm{CP}^\mathrm{th}$ &
  $ 0.09 \pm 0.19$ &
  $\mathcal{C}_\mathrm{CP}^\mathrm{exp}$ &
  $ 0.12 \pm 0.24$ \\
  $\mathcal{S}_\mathrm{CP}^\mathrm{th}$ &
  $-0.47 \pm 0.30$ &
  $\mathcal{S}_\mathrm{CP}^\mathrm{exp}$ & 
  $-0.40 \pm 0.33$ \\
  $|E_2-P_2|$ & 
  $\left\{
      \begin{array}{l}
        1.22\pm0.15 \\
        0.15 \pm 0.15
      \end{array}\right.$ & 
  $|P_2^\mathrm{GIM}-P_2|$ &
  $\left\{
      \begin{array}{l}
        0.43 \pm 0.43 \\ 
        2.87\pm0.43 
      \end{array}\right.$ \\  
  $\delta_{P}$ &
  \multicolumn{3}{l}{$\left\{
      \begin{array}{l}
        (-24 \pm 41)^\circ \\
        (-146 \pm 50)^\circ 
      \end{array} \right.$} \\ 
\end{tabular}
\end{ruledtabular}
\end{table}
As can be seen from the correlation plot in Fig.~\ref{fig:results2},
two solutions are possible, with $\vert E_2-P_2\vert$ and $\vert
P_2^\mathrm{GIM}-P_2\vert$ exchanging roles. Comparing the results of
this fit with the value for $\vert E_2-P_2\vert$ obtained from $B \to
J/\psi K^0$, it is evident that only the first solution in
Tab.~\ref{tab:results2} (with $\vert E_2-P_2\vert=1.22 \pm 0.15$) is
compatible with $SU(3)$ and with our expectations on the relative
sizes of $E_2$, $P_2$ and $P_2^\mathrm{GIM}$.  Assuming therefore that
this ambiguity is resolved in favour of the first solution, we
repeated the fit with the cut $\vert P_2^\mathrm{GIM}-P_2\vert < 2
\vert E_2-P_2\vert$. The results are presented in Fig.~\ref{fig:P2cut}
and in Tab.~\ref{tab:P2cut}. We underline the good agreement between
this result and the determination of $\vert E_2-P_2\vert$ from $B \to
J/\psi K^0$, and we conclude that there is no evidence of large
$SU(3)$-breaking effects or emission-annihilation contribution in the
determination of this parameter. We thus decide to use as input for
the determination of $\Delta \mathcal{S}$ in $B \to J/\psi K_S$ a
uniform distribution in the range $[0,2.3]$ for $\vert
P_2^\mathrm{GIM}-P_2\vert$. This corresponds to the $99.9 \%$
probability range for $\vert P_2^\mathrm{GIM}-P_2\vert$ obtained in
the fit.

\begin{table}[tb]
\caption{Results of the fit of $\bar B^0 \to J/ \psi \pi^0$ 
  with the cut $\vert P_2^\mathrm{GIM}-P_2\vert < 2 \vert
  E_2-P_2\vert$ (see the text for details).}
\label{tab:P2cut}
\begin{ruledtabular}
\begin{tabular}{cccc}
  $\mathcal{C}_\mathrm{CP}^\mathrm{th}$ &
  $ 0.09 \pm 0.19$ &
  $\mathcal{C}_\mathrm{CP}^\mathrm{exp}$ &
  $ 0.12 \pm 0.24$ \\
  $\mathcal{S}_\mathrm{CP}^\mathrm{th}$ &
  $-0.58 \pm 0.24$ &
  $\mathcal{S}_\mathrm{CP}^\mathrm{exp}$ & 
  $-0.40 \pm 0.33$ \\
  $|E_2-P_2|$ & 
        $1.22 \pm 0.15$ & 
  $|P_2^\mathrm{GIM}-P_2|$ &
        $0.38 \pm 0.38$ \\
  $\delta_{P}$ &
  \multicolumn{3}{l}{$(-34 \pm 41)^\circ \cup (-144 \pm 19)^\circ$} \\
\end{tabular}
\end{ruledtabular}
\end{table}

\begin{figure*}[tb]	
  \includegraphics[width=0.28\textwidth]{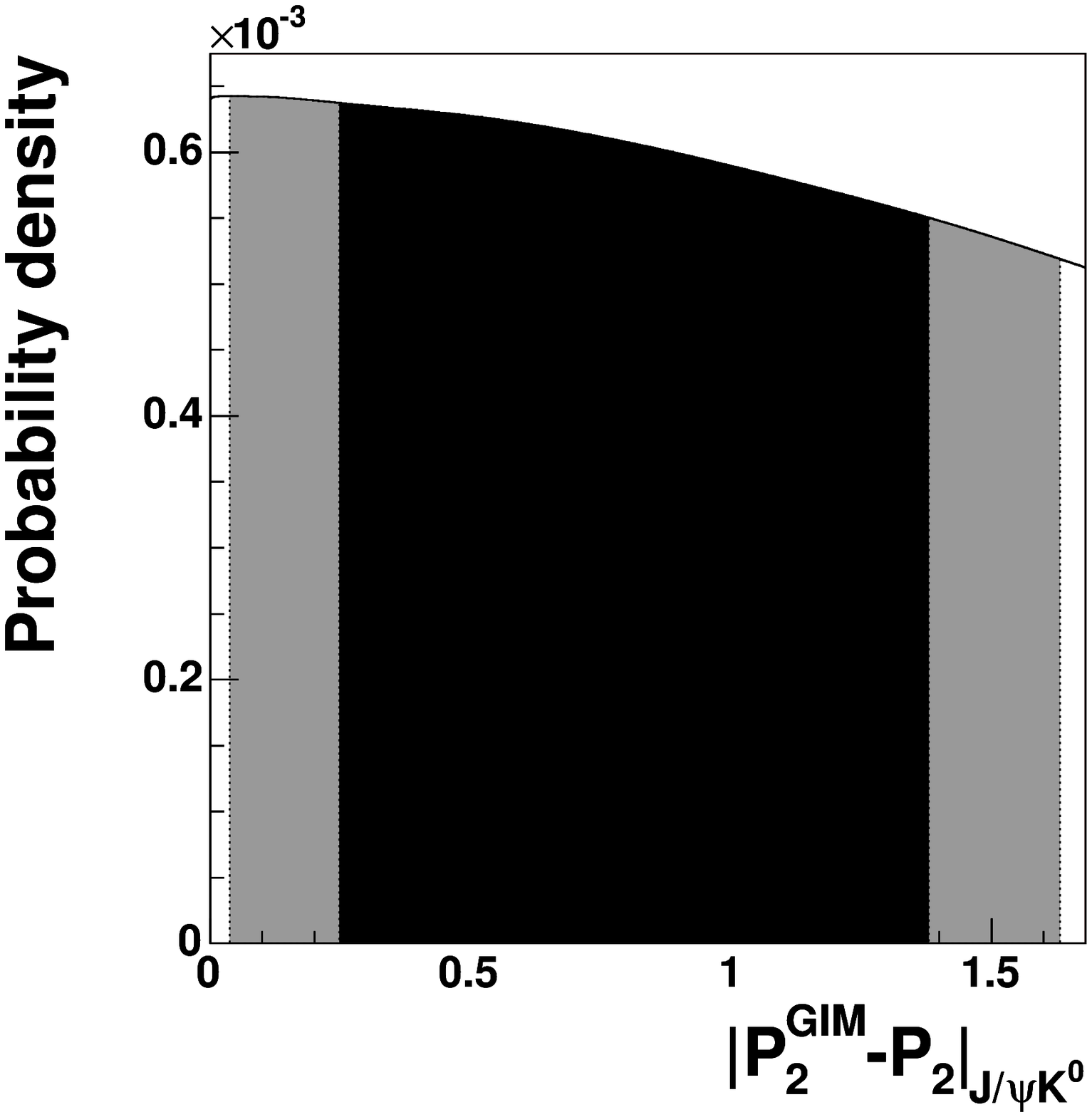} 
  \includegraphics[width=0.28\textwidth]{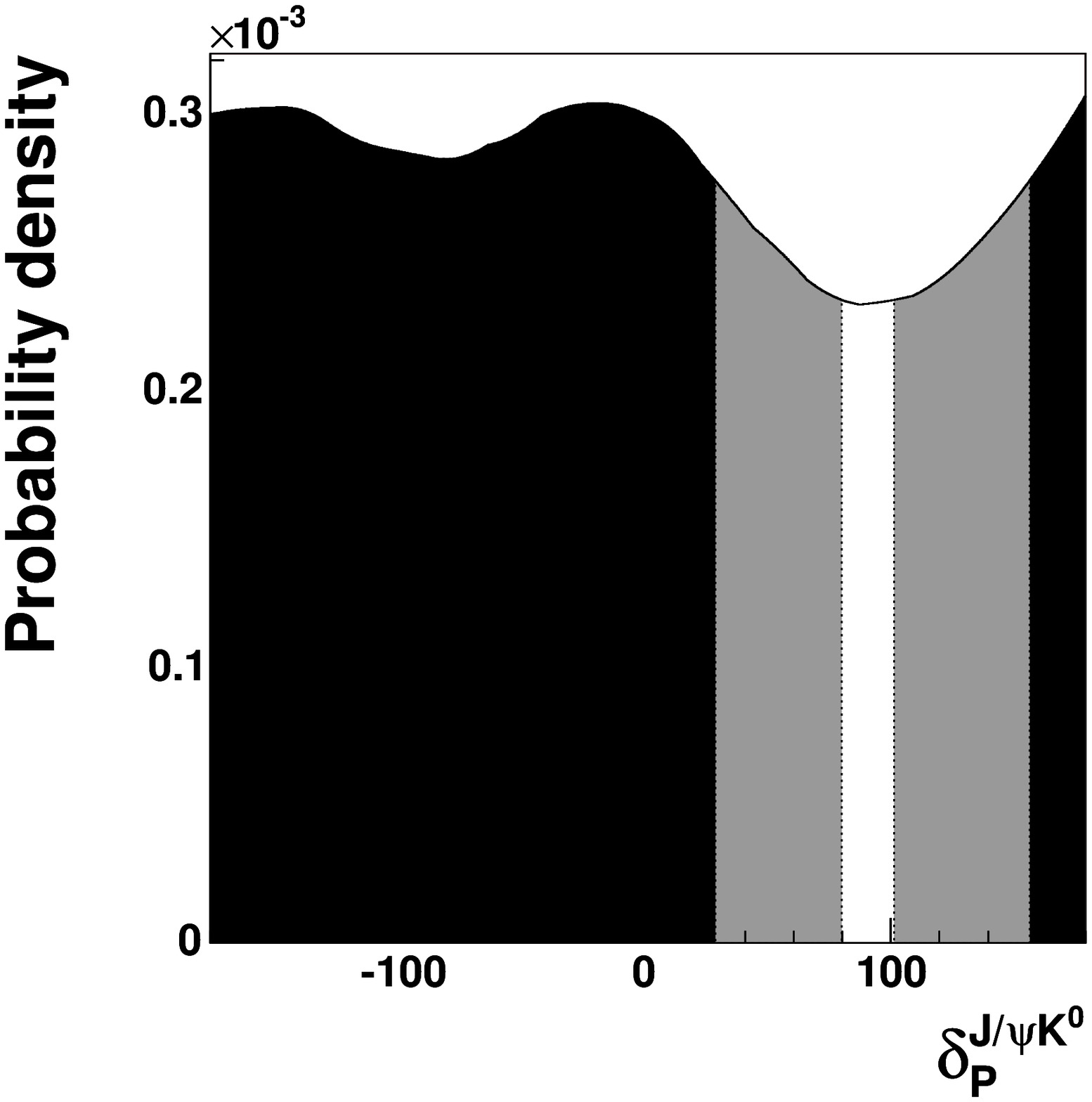} 
  \includegraphics[width=0.28\textwidth]{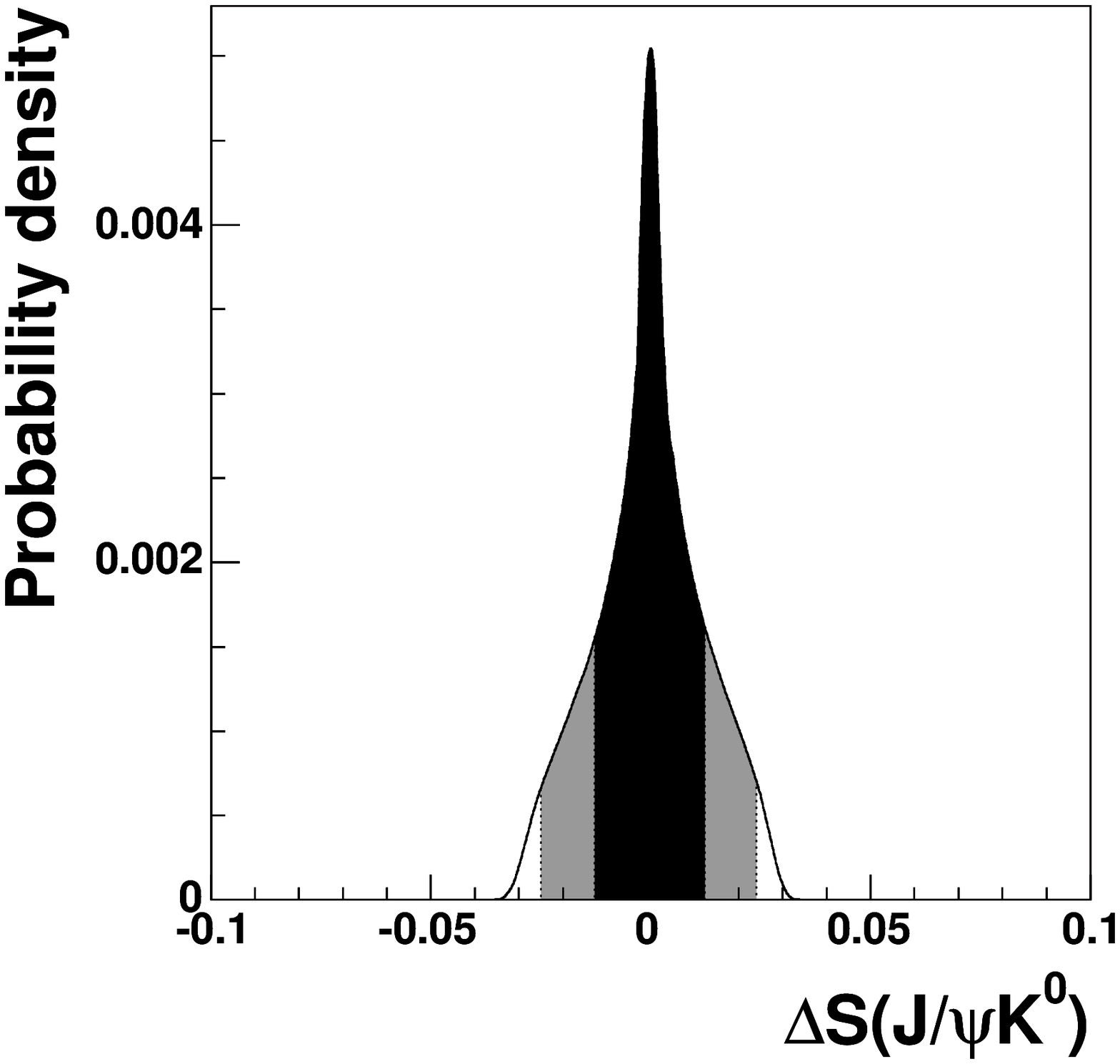}
  \caption{Output distributions of hadronic parameters
    $P_2^\mathrm{GIM}$ (left), $\delta_{P_2^\mathrm{GIM}}$ (middle)
    and $\Delta \mathcal{S}$ (right).\label{fig:results3}.}
\end{figure*}

Repeating the fit of $B^0 \to J/\psi K^0$ with the additional
contribution of $P_2^\mathrm{GIM}-P_2$ in the range obtained above, we
get the results in Tab.~\ref{tab:results3}. We also show in
Fig.~\ref{fig:results3} the output p.d.f. for $\vert
P_2^\mathrm{GIM}-P_2\vert$ and $\delta_P$, together with the
difference $\Delta \mathcal{S}$. The result is
\begin{equation}
  \label{eq:resgen}
\Delta \mathcal{S} = 0.000 \pm 0.017 \; ([-0.035,0.033]\, @ \, 95\%\;
\mathrm{prob.})\,.  
\end{equation}

Notice that, as anticipated, $\vert P_2^\mathrm{GIM}-P_2\vert$ and
$\delta_P$ are poorly determined in this fit. In particular,
Fig.~\ref{fig:results3} shows how the bound on the range of $\vert
P_2^\mathrm{GIM}-P_2\vert$ from $\bar B^0 \to J/ \psi \pi^0$ is
extremely effective in cutting out a long tail at large values of
$\vert P_2^\mathrm{GIM}-P_2\vert$, thus reducing the uncertainty on
$\Delta S$. Without this additional information, $\vert
P_2^\mathrm{GIM}-P_2\vert$ could have reached much larger values and
correspondingly we would have obtained values of $\Delta \mathcal{S}$
of order one.

\begin{table}[tb]
\caption{Results of the fit of $\bar B^0 \to J/ \psi K^0$ (see the
  text for details). $\mathcal{S}_\mathrm{CP}^\mathrm{out}$
  ($\mathcal{S}_\mathrm{CP}^\mathrm{in}$) represent the input (output)
values of $\mathcal{S}_\mathrm{CP}$ respectively.}
\label{tab:results3}
\begin{ruledtabular}
\begin{tabular}{cccc}
$\mathcal{C}_\mathrm{CP}^\mathrm{th}$ &
$0.00 \pm 0.02$ &
$\mathcal{C}_\mathrm{CP}^\mathrm{exp}$ &
$-0.01 \pm 0.04$ \\
$\mathcal{S}_\mathrm{CP}^\mathrm{out}$ &
$0.73 \pm 0.05$ &
$\mathcal{S}_\mathrm{CP}^\mathrm{in}$ &
$0.73 \pm 0.04$ \\
$|E_2-P_2|$ & 
$1.44 \pm 0.05$ & 
\multicolumn{2}{l}{$|P_2^\mathrm{GIM}-P_2|$, $\delta_{P}$: 
see text} \\
\end{tabular}
\end{ruledtabular}
\end{table}

Had we boldly borrowed from the previous step not only the range but
also the shape of $\vert P_2^\mathrm{GIM}-P_2\vert$, we would have
constrained the deviation of $\mathcal{S}_\mathrm{CP}$ from $\sin
2\beta$ even more, obtaining a value $\Delta \mathcal{S}=0.020 \pm
0.007$. However, given the theoretical uncertainties related to the
$SU(3)$ breaking and the neglected emission-annihilation contribution,
this result is quoted for illustration only, and should not be used
for phenomenology. A more reliable result can be obtained by adding a
$100\%$ error to the $SU(3)$ relation between the hadronic parameters
in the two channels. In this way we obtain
\begin{displaymath}
  \Delta \mathcal{S}=0.001 \pm 0.015\; ([-0.025,0.026]\,@\,
  95\%\;\mathrm{prob.})\,,
\end{displaymath}
fully compatible with our main result in eq.~(\ref{eq:resgen}).  
We conclude that our approach of extracting from $B \to J/\psi \pi^0$ the
\textit{range} of $\vert P_2^\mathrm{GIM}-P_2\vert$ to be used in $B
\to J/\psi K^0$ is fully consistent and does not sizably
overestimate the error in $\Delta \mathcal S$. We also stress the
importance of improving experimental results on $B \to J/\psi \pi^0$
in order to reduce the uncertainty in the extraction of $\sin 2 \beta$
from $B \to J/\psi K^0$ decays. 

Our estimate of the error in $\Delta \mathcal{S}$ is more than an
order of magnitude larger than previous estimates and comparable to
the present experimental systematic error.  This uncertainty should
therefore be included in the value and error of $\sin 2\beta$
extracted from $\mathcal{S}_\mathrm{CP}^\mathrm{exp}$.  We believe that
additional experimental information on the decay modes considered in
our analysis will allow to reduce the uncertainty in $\Delta
\mathcal{S}$ using the new method sketched in this paper and without
any need of additional theoretical input.

\begin{acknowledgments}
This work has been supported in part by Bundesministerium f\"ur
Bildung und Forschung under the contract 05HT4WOA/3 and by
the EU network ``The quest for unification" under the contract
MRTN-CT-2004-503369.
\end{acknowledgments}

\end{document}